\documentclass[reprint,aps,prl,twocolumn,amsmath,amssymb,amsfonts]{revtex4-2} %superscriptaddress,10pt,letterpaper
\usepackage[english]{babel}
\usepackage[utf8]{inputenc}
\usepackage{graphicx}
\usepackage{amsmath,amssymb,amsthm}
\usepackage{txfonts}
\usepackage{xcolor}
\usepackage{bm}
\usepackage{ulem}
\usepackage{textcomp}
\usepackage{soul}

\begin{document}

% 86mm / 178mm

\title{Supersymmetric reshaping and higher-dimensional rearrangement of photonic lattices}

\author{Tom~A.~W. Wolterink}
\email[]{tom.wolterink@uni-rostock.de}
\affiliation{Institute for Physics, University of Rostock, Albert-Einstein-Stra{\ss}e 23, 18059 Rostock, Germany}
\author{Matthias Heinrich}
\affiliation{Institute for Physics, University of Rostock, Albert-Einstein-Stra{\ss}e 23, 18059 Rostock, Germany}
\author{Alexander Szameit}
% \email[]{alexander.szameit@uni-rostock.de}
\affiliation{Institute for Physics, University of Rostock, Albert-Einstein-Stra{\ss}e 23, 18059 Rostock, Germany}

% \date{\today}
\begin{abstract}
Integrated $J_{\mathrm{x}}$ photonic lattices, inspired by the quantum harmonic oscillator and due to their equidistant eigenvalue spectrum, have been proven extremely useful for various applications, such as perfect imaging and coherent transfer of quantum states. However, to date their large-scale implementation remains challenging. We apply concepts from supersymmetry to construct two-dimensional (2D) systems with spectra identical to that of one-dimensional (1D) $J_{\mathrm{x}}$ lattices. While exhibiting different dynamics, these 2D systems retain the key imaging and state transfer properties of the 1D $J_{\mathrm{x}}$ lattice. Our method extends to other systems with separable spectra, facilitates experimental fabrication, and may increase robustness to fabrication imperfections in large-scale photonic circuits.
\end{abstract}
\maketitle

% \section{Introduction}\label{sec:introduction}
Evolution dynamics in wave-mechanical systems are governed by the full set of its modes and their respective eigenvalues. In particular, the key task of imaging light between two specific planes can be understood in this framework in ways that readily generalize from conventional free-space settings involving mirrors and lenses to waveguide-based integrated optical systems where crosstalk between the individual channels cannot be globally suppressed: imaging occurs whenever the overall phases of the individual modes simultaneously cancel except for integer multiples of $2\pi$. In discrete settings with symmetric band structures this can be enforced by actively swapping the excitations of mode pairs whose propagation constants are arranged symmetrically with respect to the center of the band \cite{Longhi2008,Szameit2008}, by continuously sweeping the wave packet through the Brillouin zone in reciprocal space by means of a transverse potential gradient (Bloch oscillations \cite{Peschel1998,Pertsch1999}), or, resonantly, via dynamic modulations \cite{Longhi2005,Longhi2006,Szameit2009}. In contrast to these active methods, image recurrence can also occur naturally as a consequence of the structure of the eigenvalue spectrum: regardless of the specific excitation, any structure with a commensurable set of eigenvalues $\{\lambda_i\}$ features periodic propagation dynamics that yield imaging after a distance corresponding to the least common multiple of the inverse eigenvalues $\{\lambda_i^{-1}\}$. This mechanism is particularly effective if the spectrum is equidistantly spaced similar to that of the harmonic oscillator. In the context of finite-size discrete systems, the so-called $J_{\mathrm{x}}$ lattice fulfills this condition and allows for various operations of relevance to classical and quantum photonic information processing, such as coherent transfer of quantum states \cite{Christandl2004,PerezLeija2013a,PerezLeija2013b}, discrete fractional Fourier transforms \cite{Weimann2016}, and the realization of saturable absorbers \cite{Teimourpour2017}. However, despite their inherently finite size, implementing large-scale $J_{\mathrm{x}}$ arrays remains experimentally challenging, as the underlying parabolic coupling profile of an $N$-site $J_{\mathrm{x}}$ array necessitates the precise realization of approximately $N/2$ different nearest-neighbor interaction strengths across a substantial dynamic range of values. To overcome these limitations, we leverage the concept of supersymmetric (SUSY) photonics \cite{Miri2013a} to design families of compact two-dimensional systems that share the spectral and imaging characteristics of $J_{\mathrm{x}}$ lattices while requiring dramatically fewer distinct couplings.

In its original context of quantum field theory, supersymmetry was developed to facilitate the theoretical description of bosons and fermions in a unified mathematical framework \cite{Ramond1971}. Subsequently, these methods have been adapted to nonrelativistic quantum mechanics \cite{Cooper1995} to systematically identify Hamiltonians with identical sets of eigenvalues. Similarly, in photonics, SUSY techniques enable the synthesis of globally phase-matched partner structures \cite{Miri2013a,Heinrich2014a} as well as the parametric design of families of isospectral \cite{Miri2014} and scattering-equivalent refractive index landscapes \cite{Heinrich2014b}. Likewise, SUSY transformations of non-Hermitian systems containing gain and loss \cite{Miri2013b} have been proposed for applications in active systems such as laser arrays \cite{Teimourpour2016a,Hokmabadi2019,Qiao2021}. While the factorization techniques typically employed in generating supersymmetric partner structures tend to be restricted to one dimension (1D), separation-of-variables methods allow them to be adapted to certain two-dimensional systems (2D) \cite{Qiao2021}, opening up a second spatial degree of freedom. Here, we present a method to construct compact 2D systems that faithfully reproduce the signature spectral and dynamic features of one-dimensional $J_{\mathrm{x}}$ arrays while dramatically reducing the number of individual structural parameters. We experimentally realize these systems in evanescently coupled waveguide arrays and systematically investigate their imaging properties. 

% \section{Transformation method}\label{sec:transformationmethod}
In conventional $J_{\mathrm{x}}$ arrays, the matrix elements of the quantum-mechanical angular momentum operator’s $x$ component are mapped onto the discrete Hamiltonian a one-dimensional array of $N$ sites \cite{Christandl2004} with a parabolic distribution of nearest-neighbor coupling coefficients $c_{n,n+1} = \frac{1}{2}\sqrt{n\left(N-n\right)}$ between sites $n$ and $n+1$. The resulting eigenvalues form an equidistant ladder of spacing 1, illustrating the close connection of this finite system to the continuous quantum-mechanical harmonic oscillator.

\begin{figure}[ht]
\includegraphics[scale=1]{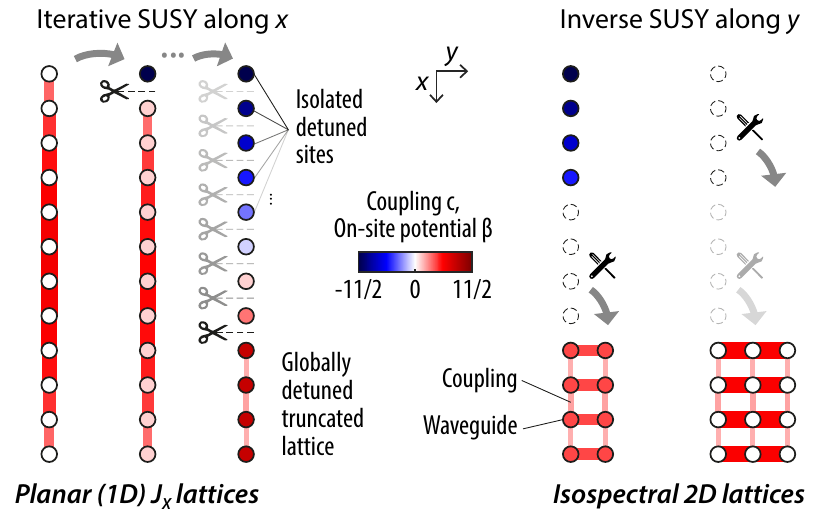}
\caption{Transforming a one-dimensional $J_{\mathrm{x}}$ array of twelve sites into a two-dimensional array through subsequent supersymmetry transformations. Color/linewidth indicates the relative strength of the coupling and detuning between sites.}
\label{fig:fig1}
\end{figure}

In order to illustrate our approach for constructing two-dimensional $J_{\mathrm{x}}$ equivalent systems, we consider the example of such a system with $N=12$ (see Fig.~\ref{fig:fig1}). Applying a discrete SUSY transformation to the Hamiltonian of the $J_{\mathrm{x}}$ array results in its first-order unbroken superpartner, which retains the spectrum of the original array with the exception of a single removed eigenvalue. In our example, the lowest eigenvalue was removed via Cholesky factorization \cite{Miri2013a}. Note that, in contrast to the QR factorization method employed to selectively remove states from of $J_{\mathrm{x}}$ lattices in \cite{Teimourpour2016b}, the Cholesky factorization relates $J_{\mathrm{x}}$ lattices with different numbers of sites, and, in our case, the superpartner turns out to be a truncated 11-site $J_{\mathrm{x}}$ array subject to a global detuning of $+\frac{1}{2}$. A single isolated site with on-site potential $\beta$ separately hosts the mode corresponding to the removed eigenvalue $-\frac{11}{2}$. Iteratively repeating this procedure on the remaining array results in a sequence of higher-order superpartners and a growing number of isolated detuned sites. Along these lines, the 8\textsuperscript{th} superpartner consists of a four-site $J_{\mathrm{x}}$ array, globally detuned by $+4$, and eight individually detuned isolated sites.

Next, we reattach these sites by a series of inverse supersymmetry transformations in the orthogonal $y$ direction, thereby constructing two-dimensional arrays. The resulting Hamiltonian can be described as the Kronecker sum $H^{(\mathrm{x})} \oplus H^{(\mathrm{y})}$ of independent components in $x$ and $y$ directions, with $H^{(\mathrm{x})}$ in our case corresponding to the Hamiltonian of the four-site $J_{\mathrm{x}}$ array detuned by $+4$ in $x$ direction and $H^{(\mathrm{y})}=0$ representing the single layer in $y$ direction. We apply inverse SUSY transformations to $H^{(\mathrm{y})}$ while keeping $H^{(\mathrm{x})}$ constant. In a first step, this reconnects the four sites of highest eigenvalue resulting in a two-layer structure, with $H^{(\mathrm{y})}$ now representing a two-site $J_{\mathrm{x}}$ array with coupling scaled by a factor of $4$ and detuned by $-2$. Repeating this procedure to reconnect the remaining sites yields a three-site $J_{\mathrm{x}}$ array with its coupling constants scaled by a factor of $4$ and a detuning of $-4$ for $H^{(\mathrm{y})}$. The final array of $4 \times 3$ sites is a two-dimensional $J_{\mathrm{x}}$ array where coupling in $y$ direction is increased fourfold compared to the $x$ direction, while the opposite detuning in $H^{(\mathrm{x})}$ and $H^{(\mathrm{y})}$ cancels out. This system resembles an anisotropic two-dimensional quantum harmonic oscillator. The newly constructed two-dimensional array shares all eigenvalues of the original one-dimensional $J_{\mathrm{x}}$ array and can be interpreted as its isospectral (broken) superpartner.

%-------------------------
%-------------------------

Note that this procedure, although detailed here for the specific case of the $J_{\mathrm{x}}$ array, holds for all systems with separable eigenvalue spectra $\Lambda=\Lambda^{(1)} \oplus \Lambda^{(2)} \oplus \cdots$ and may be also extended to higher dimensions. Moreover, by virtue of its equidistantly spaced eigenvalues, a $J_{\mathrm{x}}$ array can be transformed into different two-dimensional arrays, depending on the number $N$ of involved sites. In fact, for a $J_{\mathrm{x}}$ array of $N=N^{(x)}N^{(y)}$ sites, an entire class of isospectral two-dimensional arrays exists with Hamiltonians $H^{(x)} \oplus N^{(x)}H^{(y)}$ (or, equivalently, $N^{(y)}H^{(x)} \oplus H^{(y)}$), each representing an $N^{(x)} \times N^{(y)}$ two-dimensional $J_{\mathrm{x}}$ array whose couplings in $y$ direction have been scaled the number of sites in the $x$ direction.

From an experimental point of view, the challenge in constructing large-scale $J_{\mathrm{x}}$ arrays lies in the requirement of precisely matching approximately $N/2$ distinct coupling constants spanning a range of $c \in \left(\frac{1}{2}\sqrt{N}\ldots\frac{1}{4}N\right)$ in the limit of $N\rightarrow\infty$. The coupling values needed in the two-dimensional systems similarly fall between $c^{(x)} \in \left(\frac{1}{2}\sqrt{N^{(x)}}\ldots\frac{1}{4}N^{(x)}\right)$ in $x$ direction and $c^{(y)} \in \left(\frac{1}{2} N^{(x)}\sqrt{N^{(y)}}\ldots\frac{1}{4}N^{(x)}N^{(y)}\right)$ along $y$, respectively. While the highest coupling constant that occurs in either system is the same, our 2D $J_{\mathrm{x}}$-equivalent systems offer two advantages for implementation. First, instead of $N/2$, the nearest-neighbor couplings only take $\left(N^{(x)}+ N^{(y)}\right)/2$ distinct values, a reduction of up to $2/\sqrt{N}$. Secondly, instead of more or less homogeneously covering the entire range, the couplings of the 2D system occur in a well-separated bimodal distribution with a set of strong coupling in one of the spatial directions, and a set of weak coupling in the other. This convenient grouping allows for a more precise calibration of fabrication parameters around the respective mean values, thereby further simplifying the implementation while also allowing for potential inherent anisotropies of the experimental platform to be taken advantage of.

\begin{figure}[ht]
\includegraphics[scale=1]{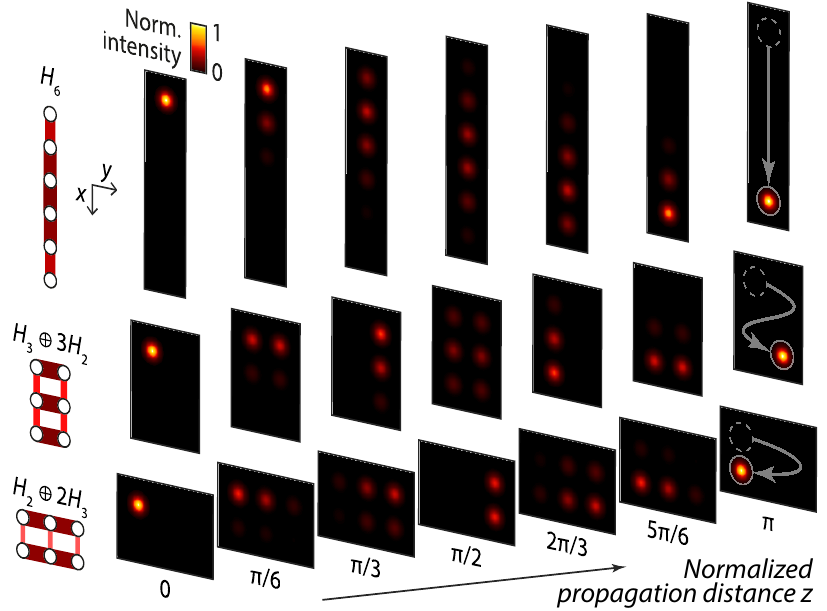}
\caption{Propagation dynamics in reshaped $J_{\mathrm{x}}$ lattices. In the schematic drawing (left), color/linewidth indicates the relative strength of the coupling between waveguides. Numerically calculated evolution (right) of an excitation of the top-left waveguide in a conventional six-waveguide $J_{\mathrm{x}}$ (top) as well as its 2D superpartners (center, bottom). In all three systems, perfect state transfer occurs at integer multiples of a normalized propagation distance of $z=\pi$.}
\label{fig:fig2}
\end{figure}

To illustrate the imaging dynamics in different isospectral systems, we consider the 2D superpartners of a six-waveguide $J_{\mathrm{x}}$ lattice and its Hamiltonian $H_6$. In line with the prime factor decomposition $6=2 \cdot 3$, this structure can either be converted into a $3 \times 2$ array described by $H_3^{(\mathrm{x})} \oplus 3H_2^{(\mathrm{y})}$, or a $2 \times 3$ array with $H_2^{(\mathrm{x})} \oplus 2H_3^{(\mathrm{y})}$, respectively. As the equidistantly spaced eigenvalue spectrum is invariant under our construction, their individual dynamics can be transformed between them via projections onto their respective eigenvectors. As a result, the imaging properties of the $J_{\mathrm{x}}$ array carry over to its superpartners, which both display perfect state transfer between opposing sites at integer multiples of the propagation distance $z=\pi$. Figure~\ref{fig:fig2} shows the numerically calculated intermediate steps of the evolution dynamics resulting from an excitation of the top-left waveguide of each of the systems. In the 2D superpartners, light independently evolves in $x$ and $y$ directions. Due to the appropriate scaling of the coupling, the evolution in $y$ is $N^{(\mathrm{x})}$ times faster than evolution in $x$. Therefore, the input wave packet is mirrored in both $x$ and $y$ directions for odd $N^{(\mathrm{x})}$, and for even $N^{(\mathrm{x})}$ only in the $x$ direction. 

%-------------------------
%-------------------------

% \section{Experimental method}\label{sec:experimentalmethod}
We experimentally probe the imaging dynamics of these systems in arrays of evanescently coupled waveguides fabricated by femtosecond-laser direct writing technique \cite{Szameit2010}. The individual guides are inscribed by focusing 274~fs laser pulses from an ultrafast fiber laser amplifier (Coherent Monaco) at a repetition rate of 333~kHz and an average power of 61~mW at a wavelength of 512~nm into fused silica (Corning 7980) through a $50\times$ microscope objective ($NA=0.6$). The glass sample is translated at a speed of 100~mm/min by means of a high-precision stage (Aerotech ALS180). Non-bridging oxygen hole centers formed during the inscription process allow for intensity dynamics of guided light to be recorded through fluorescence microscopy when excited by a helium-neon laser (633~nm). Inscribing the arrays in a slightly rotated frame of reference allows us to simultaneously observe all waveguides even in 2D lattices. In post-processing, the decrease of signal due to propagation losses is compensated by normalizing the measured intensity to the total intensity in the array at any given propagation distance. 

\begin{figure}[ht]
\includegraphics[scale=1]{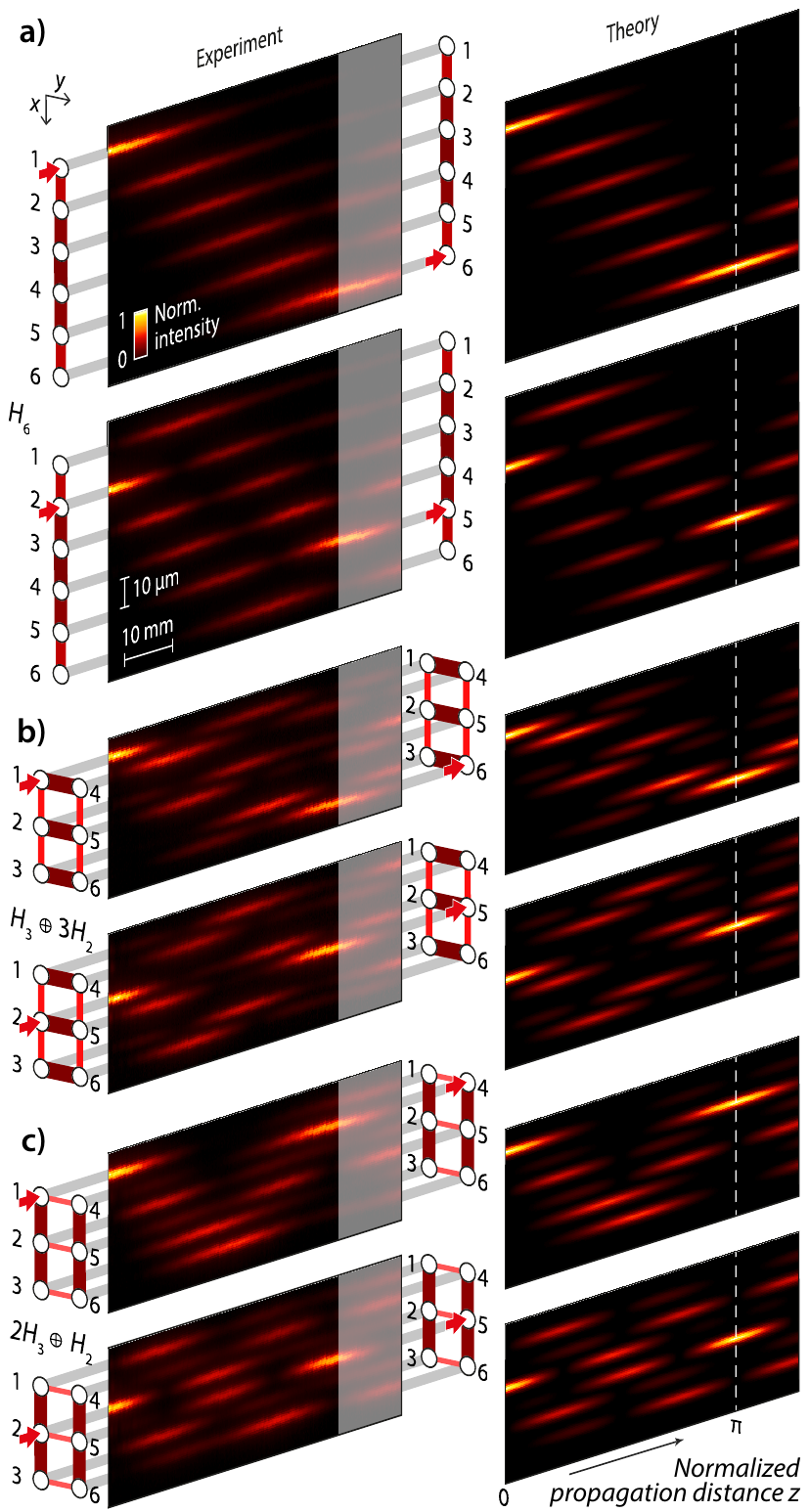}
\caption{Measured (left) and calculated (right) evolution dynamics of light in a) a six-waveguide $J_{\mathrm{x}}$ array ($H_6$) and its two-dimensional superpartner configurations b) $H_3 \oplus 3H_2$  and c) $2H_3 \oplus H_2$. In each panel, the initially excited and imaging target waveguides are marked by arrows. The imaging distance $z=\pi$ is indicated by semi-transparent overlay (experiments) and dashed lines (simulations), respectively.}
\label{fig:fig3}
\end{figure}

% \section{Results}\label{sec:results}
In a first set of experiments, we implement the conventional six-waveguide $J_{\mathrm{x}}$ array ($H_6$) and its two 2D superpartners ($H_3 \oplus 3H_2$ and $H_2 \oplus 2H_3$, respectively). The observed light evolution for two distinct input waveguides in each of these systems is shown in Fig.~\ref{fig:fig3}. In the conventional $J_{\mathrm{x}}$ array (Fig.~\ref{fig:fig3}(a)), light follows the well-known transfer trajectories that coalesce to the respective single target waveguide on the opposite side of the array at the imaging distance $z=\pi$. As the evolution continues for larger $z$, light begins to flow in the opposite direction during the second imaging period. The corresponding measurements for the two-dimensional $H_3 \oplus 3H_2$ superpartner array are shown in Fig.~\ref{fig:fig3}(b). Here, the input distribution is mirrored in both $x$ and $y$ directions. Note that, as the stronger couplings are assigned to the $y$ dimension of this structure, light here undergoes three transfers in $y$ to sync up with the single transfer in $x$ at $z=\pi$. In contrast, the evolution along $x$ in the $2H_3 \oplus H_2$ system (Fig.~\ref{fig:fig3}(c)) proceeds twice as fast as in $y$, such that the input distribution is only mirrored in the $y$ direction. Crucially, while exhibiting systematically different dynamics, each of these configurations supports the desired perfect state transfer enabled by their identical spectra. The measured dynamics are in good agreement with a theoretical coupled-mode model, illustrating that fabrication imperfections in both detuning and coupling coefficients are negligible. These effects as well as the presence of a certain amount of diagonal coupling only will begin to significantly impact the state transfer performance at larger propagation distances.

%-------------------------
%-------------------------

\begin{figure}[ht]
\includegraphics[scale=1]{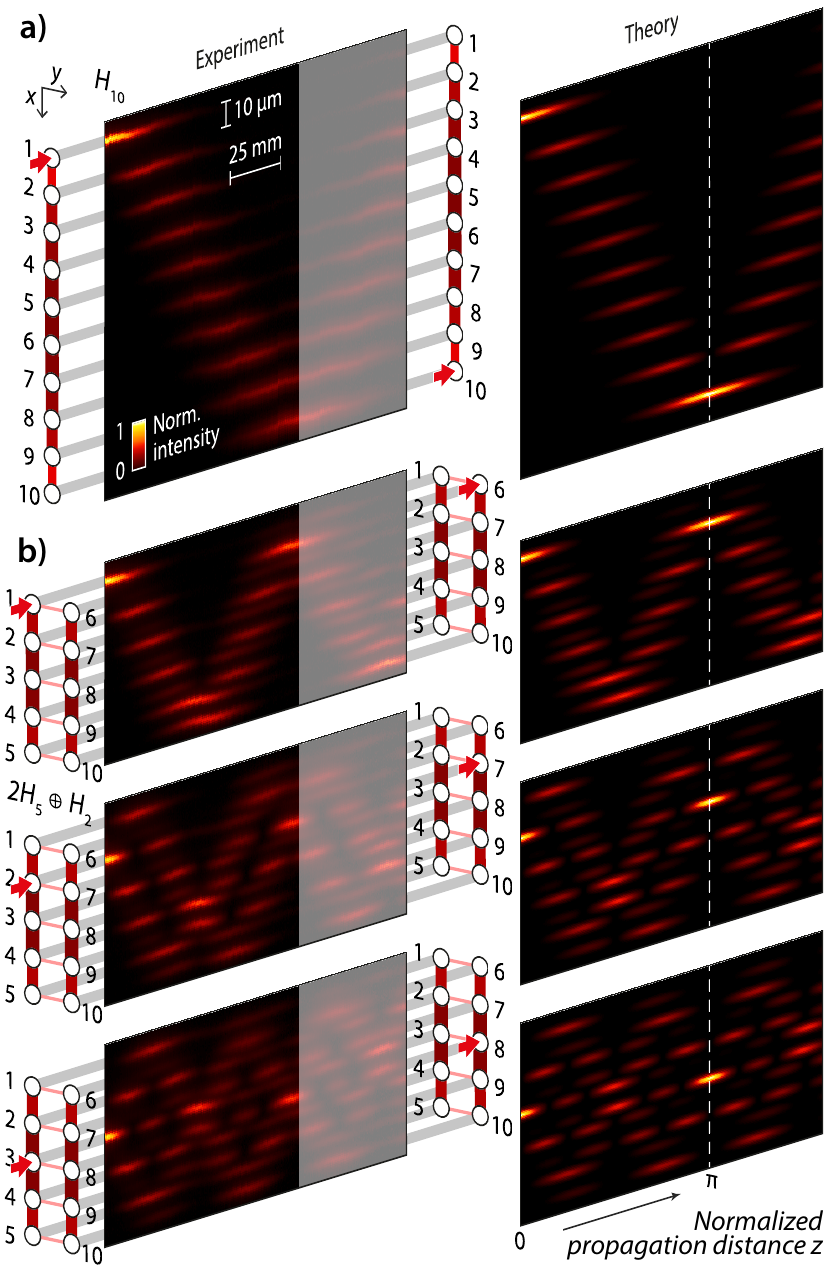}
\caption{Measured (left) and calculated (right) evolution of light in a) a ten-waveguide $J_{\mathrm{x}}$ array ($H_{10}$) and b) its 2D superpartner $2H_5 \oplus H_2$. In each panel, the initially excited and imaging target waveguides are marked by arrows. The imaging distance $z=\pi$ is indicated by semi-transparent overlay (experiments) and dashed lines (simulations), respectively.}
\label{fig:fig4}
\end{figure}

To demonstrate the versatility of our approach in more extended systems, we turn to a ten-waveguide $J_{\mathrm{x}}$ array ($H_{10}$) and its 2D counterpart, a $5\times2$ array described by $2H_5 \oplus H_2$. As shown in Fig.~\ref{fig:fig4}(a), the experimentally observed pattern deviates quite visibly from the desired perfect state transfer, with only 53\% of the light contained in the target waveguide at the imaging distance $z=\pi$. This is due to the fact that the wave packet has to sequentially traverse a substantial number of links. In contrast, the 2D superpartner is substantially more compact, and thus less susceptible to such perturbations, as can be readily seen for the three exemplary excitations in Fig.~\ref{fig:fig4}(b), where the target waveguides contain approximately 64\%, 58\% and 60\% of the light at $z=\pi$. 

%-------------------------
%-------------------------

% \section{Conclusion}\label{sec:conclusion}
In summary, we have presented a SUSY-based method to transform planar $J_{\mathrm{x}}$ lattices into isospectral two-dimensional systems. The viability of this approach was experimentally demonstrated by observing the imaging dynamics of these structures in laser-written waveguide arrays. The 2D partner arrays faithfully reproduce the hallmark features of the $J_{\mathrm{x}}$ array, such as perfect state transfer and mirroring of input field distributions, on a more compact footprint, while requiring a systematically lower number of individual structural parameters to be matched during inscription. As such, our method readily allows for increased robustness to perturbations and fabrication inaccuracies and provides the tools to maintain coherence during state transfer in large-scale photonic circuits for applications in high-dimensional quantum gates \cite{Ehrhardt2022}. Beyond the context of $J_{\mathrm{x}}$-type lattices, our method readily translates to any discrete system with a separable eigenvalue spectrum, and can be extended to higher dimensions, e.g. by leveraging the polarization degree of freedom in birefringent waveguides \cite{Ehrhardt2021}, for even higher packaging density.

\begin{acknowledgments}
The authors thank M.-A. Miri for valuable discussions. We thank C. Otto for preparing the high-quality fused-silica samples used in this work. This work was funded by Deutsche Forschungsgemeinschaft via SFB 1477 ``Light-Matter Interactions at Interfaces'', project no. 441234705. T.A.W.W. is supported by a European Commission Marie Skłodowska-Curie Actions Individual Fellowship ``Quantum correlations in PT-symmetric photonic integrated circuits'', project no. 895254.
\end{acknowledgments}

\bibliography{2DJx_ref.bib}

\end{document}